\documentclass[preprint,12pt]{elsarticle}

\usepackage{graphicx}

\usepackage{amssymb}

\journal{Computer Physics Communications}

\newcommand{\tensor}[1]{\stackrel{\leftrightarrow}{#1}}

\begin{document}

\sloppy

\begin{frontmatter}

\title{Progress in the Understanding
of the Fluctuating Lattice Boltzmann Equation}

\author[1]{Burkhard D\"unweg}
\author[1]{Ulf D. Schiller}
\author[2]{Anthony J. C. Ladd}

\address[1]{Max Planck Institute for Polymer Research,
Ackermannweg 10, D-55128 Mainz, Germany}

\address[2]{Chemical Engineering Dept.,
University of Florida,
Gainesville, FL 32611-6005, USA}

\begin{abstract}

  We give a brief account of the development of methods to include
  thermal fluctuations into lattice Boltzmann algorithms. Emphasis is
  put on our recent work (Phys. Rev. E 76, 036704 (2007)) which
  provides a clear understanding in terms of statistical mechanics.

\end{abstract}


\end{frontmatter}

The lattice Boltzmann (LB) equation has, in the last few decades,
emerged as a powerful tool to solve fluid dynamics problems
numerically \cite{sauro,benzi}. The algorithm is a fully discretized
version of the Boltzmann equation, known from the kinetic theory of
gases. Space $\vec r$ is discretized in terms of a regular (usually
simple-cubic) lattice with spacing $b$, time $t$ in terms of a time
step $h$, and velocity space in terms of a small set of velocities
$\vec c_i$ that are chosen such that $\vec c_i h$ is a vector which
connects two nearby lattice sites. For example, the popular D3Q19
model \cite{qian} employs nineteen velocities, corresponding to zero
and the six nearest and twelve next-nearest neighbors on a
simple-cubic lattice. The central quantities on which the algorithm
operates are the populations $n_i(\vec r, t)$, representing the mass
density corresponding to velocity $\vec c_i$, such that the total mass
density $\rho (\vec r, t)$ at the site $\vec r$ at time $t$ is given
by
\begin{equation}
\rho (\vec r, t) = \sum_i n_i (\vec r, t) .
\end{equation}
Similarly, the momentum density is obtained as the first velocity
moment,
\begin{equation}
\vec j (\vec r, t) = \sum_i n_i (\vec r, t) \vec c_i ,
\end{equation}
and the hydrodynamic flow velocity is given by
\begin{equation}
\vec u (\vec r, t) = \frac{\vec j (\vec r, t)}{\rho (\vec r, t)} .
\end{equation}
The algorithm is then described by the lattice Boltzmann equation
\begin{equation} \label{eq:LBE}
n_i (\vec r + \vec c_i h, t + h) =
n_i^\star (\vec r, t) =
n_i (\vec r, t) + \Delta_i \left( \{ n_i(\vec r, t) \} \right) .
\end{equation}
The collision operator $\Delta_i$ modifies the populations
on the site ($\{ n_i \}$ denotes the set of all populations on
the site), such that mass and momentum are conserved. Energy
conservation is not taken into account, since we are here
interested in an isothermal version, where the temperature
instead of the energy is fixed (formally, this corresponds to
a system with infinite heat conductivity). The conservation
equations therefore read
\begin{equation}
\sum_i \Delta_i = \sum_i \Delta_i \vec c_i = 0 .
\end{equation}
This results in a set of post-collisional populations $n_i^\star$,
which are then propagated to the neighboring sites.

In most applications, it is assumed that $\Delta_i$ is a deterministic
variable, i.~e. that it can be calculated in a unique fashion from the
populations $n_i(\vec r, t)$. This is very much in spirit of the
original continuum Boltzmann equation, and applicable to many
practical problems of fluid flow. However, for soft-matter
applications, where one is interested in Brownian motion of suspended
particles, or similar phenomena, this is not sufficient. Rather, one
must take into account that here both the lattice spacing $b$ and the
time step $h$ are so small that on these scales thermal fluctuations
are sizeable and cannot be viewed as just averaged out. Indeed,
assuming that the underlying physical model is an ideal gas, one can
see this rather easily by starting from the equation of state
\begin{equation} \label{eq:state}
k_B T = m_p c_s^2 ,
\end{equation}
where $k_B$ is Boltzmann's constant, $T$ is the absolute temperature,
$m_p$ is the mass of a gas particle, and $c_s$ is the isothermal speed
of sound ($c_s^2 = p / \rho$, where $p$ is the thermodynamic
pressure).  Usually, $c_s$ is chosen as an adjustable parameter,
picked in such a way that --- even in nonequilibrium situations like
shear flow --- the typical flow velocity $u$ is small compared to
$c_s$. This is the condition of low Mach number flow, which is needed
because of the restricted velocity space (note that $c_s$ is of the
order of the $c_i$). Furthermore, the physics of the problem usually
dictates the values of $k_B T$ and $\rho$ --- for example, we may
assume that we study water at room temperature. Equation
\ref{eq:state} then allows us to determine the mass of a gas particle,
$m_p$, which, in turn, determines the number of particles on a lattice
site (assuming a simple-cubic lattice in three dimensions),
\begin{equation} \label{eq:numparticles}
N_p = \frac{\rho b^3}{m_p} .
\end{equation}
If this number is very large, fluctuations will strongly average out,
i.~e. one can consider the single lattice site as a thermodynamic
system. This is the case for typical engineering applications.
However, if $N_p$ is comparable to unity, as it is the case for many
soft-matter applications, then fluctuations are important, and must be
taken into account in the algorithm. Since the system is an ideal gas,
$N_p$ is a random variable whose probability distribution is
Poisson. For such a distribution, the variance is identical to the
mean, i.~e. the relative importance of fluctuations is given by
\begin{equation}
Bo = \frac{ 
\left( \left< N_p^2 \right> - \left< N_p \right>^2 \right)^{1/2}}
{\left< N_p \right>} = \left< N_p \right>^{-1/2} = 
\left( \frac{m_p}{\rho b^3} \right)^{1/2} =
\left( \frac{k_B T}{\rho b^3 c_s^2} \right)^{1/2}
\end{equation}
(we coined the word ``Boltzmann number'' for this parameter). We thus
see that the degree of fluctuations is controlled by the degree of
coarse-graining, through the lattice spacing $b$. It is also useful to
introduce the parameter
\begin{equation} \label{eq:defmu}
\mu = \frac{m_p}{b^3} = \frac{k_B T}{b^3 c_s^2} ,
\end{equation}
which may be called the thermal mass density.

The question of how to actually implement these fluctuations in the
collision operator $\Delta_i$ has found different answers during the
last fifteen years, with increasing level of refinement and
understanding. In what follows, we wish to briefly outline these
developments. Since all the material has been published previously, we
would like to be brief, and refer the interested reader to the
original papers \cite{ladd1,ladd2,ronojoy,DSL} as well as to a recent
review \cite{lb_softmatter_review}, in which all the technical details
have been worked out and explained in depth.

The first implementation of a fluctuating lattice Boltzmann equation
was by Ladd \cite{ladd1,ladd2}. He started from the
well-understood deterministic version ($Bo = 0$), and added a
stochastic term $\Delta_i^\prime$ to the collision operator, with the
requirement that this is consistent, on the macroscopic scale, with
fluctuating hydrodynamics, as given by Landau and Lifshitz
\cite{landau:59}.

Let us first discuss the deterministic version in some more detail.
It is based upon a linearized collision operator,
\begin{equation}
\Delta_i = \sum_j L_{ij} (n_j - n_j^{eq}) ,
\end{equation}
where the matrix $L_{ij}$ contains constant elements, and is
implicitly given via a diagonal representation (see below), while
$n_i^{eq}$ is the lattice analog to a velocity-dependent
Maxwell-Boltzmann distribution:
\begin{equation} \label{eq:equildistrib}
n_i^{eq} \left( \rho, \vec u \right) =
a^{c_i} \rho \left( 1 + \frac{ \vec u \cdot \vec c_i }{c_s^2}
+ \frac{ \left( \vec u \cdot \vec c_i \right)^2 }{ 2 c_s^4 }
- \frac{ u^2 }{2 c_s^2} \right) .
\end{equation}
Here $c_s$ is the speed of sound, and the weights $a^{c_i} > 0$ are
normalized such that $\sum_i a^{c_i} = 1$. This notation has been
chosen in order to emphasize that, for symmetry reasons, the weights
only depend on the absolute values of the speeds $c_i$, but not on
their direction. Furthermore, the weights are adjusted in such a way
that $n_i^{eq}$ satisfies the properties
\begin{eqnarray}
\sum_i n_i^{eq} & = & \rho, \\
\sum_i n_i^{eq} \vec c_i & = & \vec j, \\
\sum_i n_i^{eq} \vec c_i \otimes \vec c_i 
& = & \rho c_s^2 \tensor 1 + \rho \vec u \otimes \vec u 
= \tensor \Pi^{eq} .
\end{eqnarray}
For D3Q19, this implies $a^{c_i} = 1/3$ for the rest population,
$a^{c_i} = 1/18$ for the nearest neighbors, and $a^{c_i} = 1/36$ for
the next-nearest neighbors. Furthermore $c_s^2 = (1/3) (b^2/h^2)$.

$L_{ij}$ is implemented as follows: First, one transforms to so-called
``modes'', i.~e. linear combinations of the $n_i$ which are adapted to
the symmetry of the problem. The first ten modes have a direct
hydrodynamic interpretation:
\begin{itemize}
  \item Mode 0: Mass density $\rho = \sum_i n_i$.
  \item Modes 1-3: Momentum density
                   $j_\alpha = \sum_i n_i c_{i \alpha}$;
                   here $\alpha$ denotes a Cartesian index.
  \item Modes 4-9: Stresses 
                   $\Pi_{\alpha \beta} = \sum_i n_i c_{i \alpha} c_{i \beta}$,
                   which are conveniently decomposed into trace and
                   traceless part:
                   $\Pi_{\alpha \beta} = 
                    \bar \Pi_{\alpha \beta} + \frac{1}{3}
                    \delta_{\alpha \beta} \Pi_{\gamma \gamma}$;
                   here we use the Einstein summation convention.
\end{itemize}
The additional modes (so-called ``kinetic'' or ``ghost'' modes) do not
have a direct relation to hydrodynamics. In the D3Q19 model, there are
nine such modes, which are explicitly listed in Ref.
\cite{lb_softmatter_review}. After having calculated the
(pre-collisional) modes, one leaves the conserved modes unchanged,
while the other modes are linearly relaxed towards their local
equilibrium value. The stresses are changed from pre- to
post-collisional values according to
\begin{eqnarray} \label{eq:det_stress_update}
  \bar \Pi^{\star \, neq}_{\alpha \beta}
  & = & \gamma_s
  \bar \Pi^{neq}_{\alpha \beta} , \\
 \nonumber
  \Pi^{\star \, neq}_{\alpha \alpha}
  & = & \gamma_b
  \Pi^{neq}_{\alpha \alpha} ,
\end{eqnarray}
where we use the notation $n_i^{neq} = n_i - n_i^{eq}$. The kinetic
modes are defined in such a way that their equilibrium part is zero,
and the action of $L_{ij}$ on them is, in the simplest version, just a
projection, such that the post-collisional kinetic modes vanish.

A Chapman-Enskog analysis shows that this procedure yields the
Navier-Stokes equations of hydrodynamics in the limit of large length
and time scales, with shear and bulk viscosities that are uniquely determined
by the values of $\gamma_s$ and $\gamma_b$, respectively.  Linear
stability requires $\left\vert \gamma_s \right\vert < 1$, $\left\vert
  \gamma_b \right\vert < 1$, corresponding to positive values of the
viscosities.

This deterministic procedure was modified by Ladd \cite{ladd1,ladd2}
by just changing Eq. \ref{eq:det_stress_update} to
\begin{eqnarray} \label{eq:stoch_stress_update}
  \bar \Pi^{\star \, neq}_{\alpha \beta}
  & = & \gamma_s
  \bar \Pi^{neq}_{\alpha \beta} + \bar R_{\alpha \beta} , \\
 \nonumber
  \Pi^{\star \, neq}_{\alpha \alpha}
  & = & \gamma_b
  \Pi^{neq}_{\alpha \alpha} + R_{\alpha \alpha} ,
\end{eqnarray}
with suitably chosen random stresses $R_{\alpha \beta}$, while the
treatment of the kinetic modes was left unchanged. The rationale
behind this procedure was that the kinetic modes do not contribute to
hydrodynamics, and the goal was to simulate the fluctuations correctly
on the hydrodynamic scale. On this scale, however, the fluctuating
stresses $\hat{R}_{\alpha \beta}$ that appear in the Navier-Stokes
equation (different from $R_{\alpha \beta}$ that appears in Eq.
\ref{eq:stoch_stress_update}) satisfy the relations \cite{landau:59}
\begin{eqnarray} \label{eq:landlifcorrelations}
  \left< \hat{R}_{\alpha \beta} \right> & = & 0 , \\
 \nonumber
  \left< \hat{R}_{\alpha \beta} \left( \vec r, t \right)
         \hat{R}_{\gamma \delta} \left( \vec r^\prime, t^\prime \right)
  \right> 
  & = &
  2 k_B T \eta_{\alpha \beta \gamma \delta} \,
  \delta \left( \vec r - \vec r^\prime \right)
  \delta \left( t - t^\prime \right) \\
  \nonumber
  & \to &
  \frac{2 k_B T}{b^3 h} \eta_{\alpha \beta \gamma \delta} \,
  \delta_{\vec r \vec r^\prime} \delta_{t t^\prime} ,
\end{eqnarray}
where $\eta_{\alpha \beta \gamma \delta}$ is the isotropic
fourth-rank viscosity tensor, parameterized by shear and
bulk viscosity, or the relaxation parameters $\gamma_s$ and $\gamma_b$.
In the last step, we have discretized the delta functions by
the lattice parameter $b$ and the time step $h$, as it is appropriate
for a lattice simulation.

One might expect that the LB noises are just given by $R_{\alpha
\beta} = \hat{R}_{\alpha \beta}$. However, this turns out not to be
correct \cite{ladd1,ladd2}. Rather, the correct fluctuating LB
stresses are obtained by a suitable modification of the amplitude. For
the shear stresses one has, for example,
\begin{equation}
\left<R_{xy}^2 \right> = (1 - \gamma_s)^2 \left< \hat{R}_{xy}^2 \right> .
\end{equation}
The same modification factor occurs for all other shear stresses, too,
while the corresponding factor for the bulk stresses is $(1 -
\gamma_b)^2$. The reason has been explained in detail in
Refs. \cite{ladd1,ladd2}; essentially the renormalization of the
amplitude comes from the fact that Eq.  \ref{eq:landlifcorrelations}
describes the physics on a more coarse-grained time scale than
Eq. \ref{eq:stoch_stress_update} --- the delta correlation in time is
in LB replaced by an exponential decay. However, the time integral of
the correlation functions must be the same in order to obtain the same
macroscopic viscosities.

Adhikari et al. \cite{ronojoy} then generalized this procedure by not
only thermalizing the stresses, but also the kinetic modes, which were
treated in a rather similar fashion to Eq. \ref{eq:stoch_stress_update}.
The argument was that the relaxation of kinetic modes introduces an
additional dissipative mechanism into the system, which should be
balanced by a compensating Langevin noise. A projection should be
viewed as the limit of such a relaxation, with relaxation parameter
$\gamma \to 0$, such that the fluctuation-dissipation relation should
hold in this case, too. While this argument makes intuitive sense, and
led to a substantially improved representation of the fluctuations at
short length scales \cite{ronojoy}, the theoretical foundation of this
procedure remained somewhat obscure (at least to the present authors).

In a recent publication \cite{DSL} we have been able to resolve these
questions by developing a first-principles theory of thermal
fluctuations in LB models. The starting point was the observation
that for a discrete system the concept of a fluctuation-dissipation
theorem should rather be replaced by the concept of detailed balance
as it applies to Monte Carlo simulations \cite{landbind}. In order to
be able to check whether an update rule satisfies or violates the
detailed-balance condition, we therefore explicitly constructed the
probability density for the random variables $n_i$ on a site in
thermal equilibrium. Taking advantage of the underlying picture of a
gas of particles, we first transform from the $n_i$ to variables
$\nu_i$, the number of particles on the site which have velocity $\vec
c_i$ (cf. Eqs. \ref{eq:numparticles} and \ref{eq:defmu}):
\begin{equation}
\nu_i = \frac{n_i}{\mu} .
\end{equation}
In terms of these variables, the probability density (except for
normalization, which is unimportant for our purposes) is written as
\begin{equation} \label{eq:probabildist}
  P \left( \left\{ \nu_i \right\} \right) \propto
  \left( \prod_i
  \frac{ \bar \nu_i^{\nu_i} }{\nu_i !}
  \exp \left( - \bar \nu_i \right) \right)
  \delta \left( \sum_i \mu \nu_i - \rho \right)
  \delta \left( \sum_i \mu \vec c_i \nu_i - \vec j \right) .
\end{equation}
The underlying picture is that of a ``velocity bin'' $i$ in thermal
contact with a huge reservoir of particles, resulting in a Poisson
distribution of the variable $\nu_i$. This distribution is
characterized by its mean value $\bar \nu_i$, which, for reasons of
consistency with the deterministic version, should be proportional to
the weight $a^{c_i}$ (see Eq. \ref{eq:equildistrib}). Normalization
requires
\begin{equation}
\bar \nu_i = \frac{a^{c_i} \rho}{\mu} .
\end{equation}
Equation \ref{eq:probabildist} then results from assuming that all the
velocity bins on the site are statistically independent, except for
the constraints of conserved mass and momentum, which are taken into
account by the delta functions, in close analogy to the statistical
description of the microcanonical ensemble \cite{landau:statphys}.

The further development is somewhat technical but straightforward and
shall be sketched only briefly. We use Stirling's formula and
transform back to the $n_i$ to write the factor in front of the delta
functions as $\exp(S)$, where the entropy $S$ has a Boltzmann-like
form. Maximizing $P$ is equivalent to maximizing $S$ under the
constraints of given values for $\rho$ and $\vec j$, and the solution
of this problem, up to second order in $u$, is just
Eq.~\ref{eq:equildistrib}, as is well-known from previous studies of
the ``entropic lattice Boltzmann'' approach \cite{karlin}.
Fluctuations around the most probable populations are described by
$n_i^{neq}$, which, within a saddle-point approximation, obey a
Gaussian distribution, whose variance is, within a $u \to 0$
approximation, given by $\mu \rho a^{c_i}$. Normalizing the
fluctuations to unit variance, followed by an orthonormal
transformation to normalized modes $\hat m_k^{neq}$, yields a very
simple form for the probability distribution,
\begin{equation}
P \left( \{ \hat{m_k}^{neq} \} \right)
\propto \exp \left( - \frac{1}{2} \sum_{k > 3}
\hat{m}_k^{neq \, 2} \right) ,
\end{equation}
where modes $i = 0, \ldots, 3$ do not occur due to mass and momentum
conservation. These modes are updated according to the rule
\begin{equation}
\hat{m}_k^{\star neq} = \gamma_k \hat{m}_k^{neq} + \varphi_k r_k ,
\end{equation}
with adjustable parameters $\gamma_k$, $\varphi_k$, and normalized,
independent Gaussian random numbers $r_k$. It is then straightforward
to show \cite{DSL,lb_softmatter_review} that detailed balance holds
exactly for
\begin{equation} \label{eq:detailedbalance3}
\varphi_k = \left( 1 - \gamma_k^2 \right)^{1/2} ,
\end{equation}
which turns out to be identical to the prescription of Adhikari et al.
\cite{ronojoy}. This shows that the stochastic analog of projecting
out the kinetic modes is to sample them from scratch, and explains the
non-trivial prefactor in the fluctuating stresses in a straightforward
way. Furthermore \cite{DSL,lb_softmatter_review}, one may apply the
Chapman-Enskog procedure to the stochastic version of the
algorithm. This shows in a particularly concise way that the behavior
in the hydrodynamic limit is given by Landau-Lifshitz fluctuating
hydrodynamics \cite{landau:59}, and that the details of the dynamics
of the kinetic modes are indeed immaterial for the behavior in that
limit, as already anticipated in Refs. \cite{ladd1,ladd2}. For
practical simulations, however, one should prefer the more recent
version which does satisfy detailed balance on the local scale as
well. We believe that this is really an improvement that outweighs
the computational costs, which are unfortunately not completely
negligible. While simple LB algorithms have so few operations per
collision step that they are typically limited by the bandwith of
memory access in the streaming step \cite{wellein}, this does not seem
to be true here, where the generation of random numbers combined with
the linear transformation to mode space and back contributes
noticeably. In practice, one may say that the additional
thermalization of the kinetic modes will slow down the algorithm by
roughly $20 \% \ldots 40 \%$ --- at least this is what we observed for
our D3Q19 implementation, see Table \ref{tab:perf_data}. For large
lattices the memory bottlenecks become more important than for small
ones; for this reason, the simulations become systematically slower,
while the performance difference between ``stresses-only'' vs.
full thermalization becomes less pronounced.

\begin{table}
\begin{center}
\begin{tabular}{| c | c | c |} \hline
$L$ & stresses--only & full thermalization \\
\hline
10  &   0.74         &  0.47 \\
20  &   0.66         &  0.44 \\
30  &   0.52         &  0.39 \\
\hline
\end{tabular}
\end{center}
\caption{Performance of the stochastic D3Q19 algorithm, using an
  implementation on a 64-bit AMD Athlon 3500+ processor
  with 2.2 GHz CPU speed and 512 kB cache size. The program is
  part of the Mainz ESPResSo \cite{espresso} package. Simulations
  were run on simple-cubic lattices of size $L^3$ for $10^5$ lattice
  sweeps, and Gaussian random numbers were generated by the Box-Muller
  \cite{numrecip} method. Performance data are given in MLUPS
  (million lattice-site updates per second). 
}
\label{tab:perf_data}
\end{table}

So far, only the case of an isothermal ideal gas has been thoroughly
understood. For the future, it is hoped that the present theoretical
approach will also help develop an improved understanding of systems
with non-trivial equations of state, and systems where thermal
conduction and energy conservation are taken into account.


\begin{thebibliography}{10}

\bibitem{sauro}
Succi, S.,
\newblock {\em The Lattice Boltzmann Equation for Fluid Dynamics and Beyond},
\newblock Oxford University Press, Oxford, 2001.

\bibitem{benzi}
Benzi, R., Succi, S., and Vergassola, M.,
\newblock Phys. Rep. {\bf 222} (1992) 145.

\bibitem{qian}
Qian, Y.~H., D'Humieres, D., and Lallemand, P.,
\newblock Europhys. Lett. {\bf 17} (1992) 479.

\bibitem{ladd1}
Ladd, A. J.~C.,
\newblock J. Fluid Mech. {\bf 271} (1994) 285.

\bibitem{ladd2}
Ladd, A. J.~C.,
\newblock J. Fluid Mech. {\bf 271} (1994) 311.

\bibitem{ronojoy}
Adhikari, R., Stratford, K., Cates, M.~E., and Wagner, A.~J.,
\newblock Europhys. Lett. {\bf 71} (2005) 473.

\bibitem{DSL}
D{\"u}nweg, B., Schiller, U.~D., and Ladd, A. J.~C.,
\newblock Phys. Rev. E {\bf 76} (2007) 036704.

\bibitem{lb_softmatter_review}
D{\"u}nweg, B. and Ladd, A. J.~C.,
\newblock Adv. Polym. Sci. {\bf 221} (2009) 89.

\bibitem{landau:59}
Landau, L.~D. and Lifshitz, E.~M.,
\newblock {\em Fluid Mechanics},
\newblock Addi\-son-Wesley, Reading, 1959.

\bibitem{landbind}
Landau, D.~P. and Binder, K.,
\newblock {\em A Guide to Monte Carlo Simulations in Statistical Physics},
\newblock Cambridge University Press, Cambridge, 2000.

\bibitem{landau:statphys}
Landau, L.~D. and Lifshitz, E.~M.,
\newblock {\em Statistical Physics},
\newblock Addi\-son-Wesley, Reading, 1969.

\bibitem{karlin}
Karlin, I.~V., Ferrante, A., and \"Ottinger, H.~C.,
\newblock Europhys. Lett. {\bf 47} (1999) 182.

\bibitem{wellein}
Wellein, G., Zeiser, T., Hager, G., and Donath, S.,
\newblock Computers \& Fluids {\bf 35} (2006) 910.

\bibitem{espresso}
Limbach, H.-J., Arnold, A., Mann, B. A., and Holm, C.,
\newblock Comput. Phys. Commun. {\bf 174} (2006) 704.

\bibitem{numrecip}
Press, W. H., Flannery, B. P., Teukolsky, S. A., and Vetterling, W. T.,
\newblock {\em Numerical Recipes},
\newblock Cambridge University Press, Cambridge, 1986.

\end{thebibliography}
\end{document}